\documentclass[a4paper]{JHEP3}
\usepackage{epsfig}
\usepackage{graphicx}
\usepackage{amssymb}
\usepackage{latexsym}

\newcommand{\be}{\begin{equation}}
\newcommand{\ee}{\end{equation}}
\newcommand{\bea}{\begin{eqnarray}}
\newcommand{\eea}{\end{eqnarray}}

\def\Pcm#1{{\mathcal{#1}}}
\newcommand{\del}{\partial}
\newcommand{\td}{\textup{d}}  
\def\eqref#1{(\ref{#1})}
\def\er#1{eqn.\eqref{#1}}
\def\nn{\nonumber}

\title{Spectrum to all orders of Polchinski-Strominger {Effective} String Theory of  Polyakov-Liouville Type}
\author{N.D. Hari Dass
\\ CHEP , Indian Institute of Science, Bangalore 560012, India \footnote{DAE Raja Ramanna Fellow} \\
Poornaprajna Institute of Scientific Research, Bangalore 560080, India\\
 Email: \email{dass@cts.iisc.ernet.in}}
\author{Peter Matlock
\\ Poornaprajna Institute of Scientific Research, Bangalore 560080, India\\
 Email: \email{lateflip@gmail.com}}
\author{Yashas Bharadwaj
\\ CHEP , Indian Institute of Science, Bangalore 560012, India \\
 Email: \email{bharadwajy@cts.iisc.ernet.in}}

\abstract{
The spectrum of a Polchinski-Strominger type effective string theory extended to all orders herein
called an effective string theory of the \emph{Polyakov-Liouville Type} (for obvious reasons), is investigated to all orders
in the small parameter $R^{-1}$. Here $2\pi R$ is the length of the \emph{closed} string. It is established that to
\emph{all orders} the spectrum of this theory is \emph{identical} to that of the free bosonic string theory. While
the latter is consistent only in the critical dimension $D_c=26$, the PS-
type effective string theories are by
construction consistent in \emph{all} dimensions. This work extends earlier results by Drummond, and, by Hari Dass
and Matlock to order $R^{-3}$. When combined with Drummond's results about absence of candidate actions at orders
$R^{-4},R^{-5}$, our results imply that the spectrum of \emph{all} effective string theories coincides with that
of free bosonic string theories to order $R^{-5}$. This agrees in part with some recent results by Aharony and Karzbrun. 
Our work is the first all order analysis of any effective string theory.
}


\begin{document}

\section{Introduction}
\label{sec_intro}
String-like defects or solitons occur in a wide variety of physical systems. Some well-known examples are vortices
in superfluids, the Nielsen-Olesen vortices of quantum field theories, vortices in Bose-Einstein condensates and
QCD-strings(for reviews see \cite{bali,kutilat05}). Depending on the particular physical circumstances, these objects can behave quantum-mechanically. The
challenge then is to find consistent ways of describing string-like objects quantum mechanically in \emph{arbitrary}
dimensions. It should be recalled that fundamental string theories are consistent only in the so called critical
dimensions. Polyakov \cite{polya} gave formulations of string theories in \emph{sub-critical} dimensions; 
his ideas play a central role in what follows.
It would be an overkill, and in all likelihood even incorrect, to treat the above mentioned string-like defects
as fundamental strings. A more pragmatic approach would be to treat them in some effective manner much as interactions
of pions are so succesfully described in terms of effective field theories without any pretenses about these field theories
being fundamental at all scales.

Two such approaches to effective string theories exist in the literature. One due to L\"uscher and collaborators \cite{lwearly}, is formulated
entirely in terms of the $D-2$ \emph{transverse} degrees of freedom. It is a case where the \emph{gauge} is fixed completely
without any \emph{residual} invariance left. This approach was further developed in \cite{others}. The recent work of Aharony and 
Karzbrun \cite{aharony} has followed this approach in addressing
the issue of spectrum of effective string theories to higher orders. We \cite{ouruniv,field,covariant}, along with Drummond \cite{drumorig,drumrep}, have on the other hand followed
the approach pioneered by Polchinski and Strominger \cite{ps}. In the latter approach, the theories are invariant under \emph{conformal
transformations} and the physical states are obtained by requiring that the generators of conformal transformations
annihilate them. These too are gauge-fixed theories but with leftover residual invariances characterized by conformal
transformations. It is worth emphasizing that the physical basis of both approaches is that the degrees of freedom are 
transverse.

The organisation of this paper is as follows: in section \ref{sec_PS} we give a brief introduction to the main features of the
PS-formalism; in section \ref{sec_absence} we briefly summarise earlier work proving the absence of the $R^{-3}$ corrections to the
spectrum of the effective theories in comparison to that of the free bosonic string theory \cite{drumorig,ouruniv,drumrep,rep2rep,field}. We then show in section \ref{sec_all_analysis} 
how to extend
to all orders the particular action that PS elaborated in their original work. In section \ref{sec_liou_all} we compute the equations of motion 
(EOM) and the stress tensor of this extended theory to all orders. As an intermediate step to our final result we establish in section
\ref{sec_propT} several important properties of the class of stress tensors relevant to this paper; there we also establish some facts about field redefinitions.
The field redefinitions of interest are shown to be benign in the sense that there are no problems of
operator orderings usually encountered with generic field redefinitions. Issues involved with generic field redefinitions
have been elaborated in \cite{field}. 
Finally, in section \ref{spectrumall} we prove the absence of corrections to all orders in comparison to the results of the free bosonic 
theory . We conclude, in section \ref{sec_discussion}, by
comparing our results with those recently obtained by Aharony and Karzbrun \cite{aharony}.

\section{Polchinski-Strominger Theory}
\label{sec_PS}
Here, we review the analysis given by Polchinski and Strominger \cite{ps}. In what follows, we distinguish the field definition that
they use by \emph{primed} quantities.

They begin with the action
\begin{equation}
\label{psaction}
S_{PS} = S_0+\frac{\beta}{4\pi} \int \td\tau^+ \td\tau^- \bigg\{
\frac{\del_+^2 X^\prime\cdot\del_- X^\prime \del_+ X^\prime\cdot\del_-^2 X^\prime}{(\del_+X^\prime\cdot\del_-X^\prime)^2} 
+\Pcm{O}(R^{-3})
\bigg\}
.\end{equation}
where
\begin{equation}
S_0 = \frac{1}{4\pi a^2} \int \td\tau^+ \td\tau^- L^\prime
 = \frac{1}{4\pi a^2} \int \td\tau^+ \td\tau^- 
  \del_+ {X^\mu}^\prime \del_- {X_\mu} ^\prime
\end{equation}
is the action for the free bosonic string theory and is consistent quantum mechanically only in $D=26$ space-time dimensions.
The action $S_{PS}$ is invariant under the modified conformal transformations
\begin{equation}
\label{modtrans}
\delta_- {X^\mu}^\prime = \epsilon^-(\tau^-)\del_- {X^\mu}^\prime - \frac{\beta a^2}{2}\del_-^2 \epsilon^-(\tau^-)
\frac{\del_+ {X^\mu}^\prime}{\del_+X^\prime\cdot\del_-X^\prime}
,\end{equation}
(and another, $\delta_+X^\prime$, with $+$ and $-$ interchanged)
leading to the energy momentum tensor (which agrees with eqn(11) of \cite{ps}
to the relevant order)
\begin{eqnarray}
\label{fullT__}
{T^\prime_{--}}^{\textup{PS}} &=& -\frac{1}{2a^2}\del_-X^\prime\cdot\del_-X^\prime
+\frac{\beta}{2L^2}\big(
     L^\prime\del_-^2 L^\prime-(\del_- L^\prime)^2 \nn\\
&+&\del_-X^\prime\cdot\del_-X^\prime \del_+^2 X^\prime\cdot\del_-^2 X^\prime -\del_+ L \del_-X^\prime\cdot\del_-^2 X^\prime 
     \big)
\end{eqnarray}
where we have omitted terms proportional to the leading-order equation 
of motion,
$ \del_+\del_- {X^\mu}^\prime = 0 $ which has a solution
${X^\mu}^\prime_{\textup{cl}} = e^\mu_+R\tau^+ + e^\mu_- R \tau^-$;
here $e_-^2=e_+^2=0$ and $e_+\cdot e_- = -1/2$, as in \cite{ps}.
Fluctuations around the classical solution are denoted by ${Y^\mu}^\prime$, so that
$
{X^\mu}^\prime = {X^\mu}^\prime_{\textup{cl}}+{Y^\mu}^\prime
$.
The energy-momentum tensor in terms of the fluctuation field is then
\begin{equation}
\label{T_2}
T^\prime_{--} = -\frac{R}{a^2}e_\cdot\del_-Y^\prime -\frac{1}{2a^2}\del_-Y^\prime\cdot\del_-Y^\prime
-\frac{\beta}{R}e_+\cdot\del_-^3 Y^\prime+\ldots
\end{equation}
Using standard techniques the spectrum of this effective theory was 
worked out by PS at the leading order.  We
briefly reproduce their results here in order to set the stage for the
rest of the paper. The oscillator representation for the Virasoro generators, 
provided by 
$\del_-{Y^\mu}^\prime = a\sum_{m=-\infty}^{\infty}{\alpha_m^\mu}^\prime e^{-im\tau^-}$, 
is given by
\begin{eqnarray}
\label{virasoro}
L_n^\prime &=& \frac{R}{a}e_-\cdot\alpha_n^\prime
+\frac12\sum_{m=-\infty}^\infty :\alpha_{n-m}^\prime\cdot\alpha_m^\prime: \nn\\
&+&\frac{\beta}{2}\delta_{n}
-\frac{a\beta n^2}{R}e_+\cdot\alpha_n^\prime + \Pcm{O}(R^{-2})
.\end{eqnarray}
It is easily verified on using the canonical commutation relations
\begin{equation}
\label{ccr}
[{\alpha_m^\mu}^\prime,{\alpha_n^\nu}^\prime] = m\eta^{\mu\nu}~\delta_{m+n,0}
.\end{equation}
that these generators satisfy
\begin{equation}
\label{viralg}
[L_m^\prime,L_n^\prime] = (m-n)L_{m+n}^\prime+\frac{(D+12\beta)}{12}(m^3-m)
.\end{equation}
One must add the contribution $-26$ from the
ghosts, leading to the total central charge $D+12\beta-26$. 
Vanishing
of the conformal anomaly thus requires
\begin{equation}
\label{betacrit}
\beta_c = -\frac{D-26}{12}
,\end{equation}
valid for any dimension $D$.

The quantum ground state is $|k,k;0\rangle$ which is also an
eigenstate of $\alpha_0^\mu$ and ${\tilde\alpha}_0^\mu$ with common
eigenvalue $ak^\mu$.  This state is annihilated by all $\alpha_n^\mu$
for positive-definite $n$.  The ground state momentum is
$p^\mu_{\textup{gnd}} = \frac{R}{2a^2}(e_+^\mu+e_-^\mu) + k^\mu$ while
the total rest energy is
\begin{equation}
\label{restenergy}
(-p^2)^{1/2} = \sqrt{
\left(\frac{R}{2a^2}\right)^2
-k^2-\frac{R}{a^2}(e_++e_-)\cdot k } 
.\end{equation}
The physical state conditions $L_0^\prime = {\tilde{L}}_0^\prime = 1$ fix $k$, so that
\begin{equation}
\label{leadingvir}
k^1 = 0,\qquad
k^2+\frac{R}{a^2}(e_++e_-)\cdot k+\frac{2N}{a^2} = \frac{(2-\beta_c)}{a^2}
.\end{equation}
The first follows from the periodic boundary condition for the closed string
which gives $e_+^\mu-e_-^\mu = \delta^\mu_1$. Substituting the critical value
$\beta_c = (26-D)/12$ one arrives at
\begin{equation}
(-p^2)^{1/2} = \frac{R}{2a^2}\sqrt{1-\{\frac{D-2}{12}-2N\}\left(\frac{2a}{R}\right)^2}
,\end{equation}
which is the precise analog of the result obtained by Arvis for open strings
\cite{Arv}. Here $N$ is the occupation number for the excited state.

Expanding this for the ground state and keeping
only the first correction, one obtains for the static potential
\begin{equation}
V(r) = \frac{R}{2a^2} -\frac{D-2}{12~R}+\cdots
.\end{equation}
\section{Absence of corrections to the spectrum at $R^{-3}$ level}
\label{sec_absence}
\subsection{Absence of candidate actions at $R^{-3}$}
\label{sec_noaction}
It was already stated in \cite{ps} that the next candidate action term is
of order $R^{-4}$. This being  central to the issue of the spectrum, a general proof is in order.
Such a proof was given by \cite{drumorig,ouruniv,drumrep,rep2rep,field}. 
The proofs were based on a combinatoric analysis of the PS-algorithm for constructing effective actions.
We shall not give the details here.The final result was that there are no candidate actions at 
$R^{-3},R^{-4},R^{-5}$ levels. The most robust of these conclusions is the absence of candidate actions at $R^{-3}$
level. The proof of absence of actions at $R^{-4},R^{-5}$ involved making several field redefinitions whose 
reliability would need further analysis. Various issues arising out of field redefinitions were analysed in
\cite{field}.

\subsection{Absence of $R^{-3}$ corrections to the spectrum}
\label{sec_rminus3}
The work described here \cite{drumorig,ouruniv,drumrep,rep2rep,field}(see also \cite{unpub}) was motivated by the high accuracy numerical simulations
\cite{ournum} which were based on the remarkable developments by L\"uscher and Weisz \cite{lwrecent}. All higher corrections are determined by
$k^2+\frac{R}{a^2}(e_++e_-)\cdot k$ (\er{restenergy}) which was 
calculated to leading order in \er{leadingvir}. An order-$R^{-n}$
correction to this results in order-$R^{-n-1}$ and higher
corrections to the spectrum. 
As this is just a sum of the $L_0^\prime$
and ${\tilde{L}}_0^\prime$ conditions, the latter(equivalently $T_{--}^\prime$)
need to be calculated to
order-$R^{-2}$ to reach conclusions about the spectrum at $R^{-3}$ level.

The transformation laws \eqref{modtrans} have a leading part linear
in $R$; therefore, $R^{-3}$ terms in the action 
induce $R^{-2}$ corrections to $T_{--}^\prime$. 
This
is why the absence of candidate actions at $R^{-3}$-level needed to be
established so carefully. Absence of such actions also means that 
$T_{--}^\prime$ in \er{fullT__} 
is actually valid 
to order-$R^{-2}$. We give here the 
expression to the desired order, modulo terms proportional to leading order EOM;
\begin{eqnarray}
\label{highT}
&&T_{--}^\prime= -\frac{R}{a^2}e_-\cdot\del_-Y^\prime-\frac{1}{2a^2}\del_-Y^\prime
\cdot\del_-Y^\prime-\frac{\beta}{R}e_+\cdot\del_-^3Y^\prime \nn\\
&&-\frac{\beta}{R^2}\big[2(e_+\cdot\del_-^2Y^\prime)^2
+2e_+\cdot\del_-^3Y^\prime(e_+\cdot\del_-Y^\prime+e_-\cdot\del_+Y^\prime) \nn\\
&&+2e_-\cdot\del_-^2Y^\prime e_-\cdot\del_+^2Y^\prime+\del_+Y^\prime\cdot\del_-^3Y^\prime
\big]
.\end{eqnarray}

We now point to an issue with this equation; this is a recurring issue with all effective actions and will
be seen to play an important role later in this paper too. The $T_{--}^\prime$ of \er{highT} does not seem \emph{holomorphic} 
because of the $+$-derivative terms occurring in it.
But $T_{--}^\prime$ satisfies $\del_+ T_{--}^\prime=0$ \emph{on-shell}. 
The resolution lies in the fact that the solution of the
full equation of motion can no longer be split into a sum of
holomorphic and antiholomorphic pieces \cite{ouruniv}.

The relevant EOM to order $R^{-3}$ is given by
\begin{eqnarray}
\label{R3EOM}
&&\frac{2}{a^2}\del_{+-}{Y^\mu}^\prime =
- 4\frac{\beta}{R^2}\del_+^2\del_-^2 {Y^\mu}^\prime 
-4\frac{\beta}{R^3}\bigg[
\del_+^2\big\{
              \del_-^2{Y\mu}^\prime(e_+\cdot\del_-Y^\prime+e_-\cdot\del_+Y^\prime)
                \big\} \nn\\
        &&+\del_-^2\big\{
               \del_+^2{Y^\mu}^\prime(e_+\cdot\del_-Y^\prime+e_-\cdot\del_+Y^\prime)
                 \big\}  
 + 4 e_+^\mu\del_-(\del_+^2\cdot\del_-^2Y^\prime)+e_-^\mu\del_+(\del_+^2Y^\prime\cdot\del_-^2Y^\prime)
\bigg]
.\end{eqnarray}
We can solve this equation iteratively by writing ${Y^\mu}^\prime = {Y^\mu}^\prime_0+{Y^\mu}^\prime_1$ 
where ${Y^\mu}^\prime_0$ is a solution of the equation of motion of the free theory.
The result is
\begin{equation}
\frac{2}{a^2}\del_{-}{Y^\prime_1}^\mu = 4\frac{\beta}{R^3}
\big( e_+^\mu\del_+Y_0^\prime\cdot\del_-^3Y_0^\prime+e_-^\mu\del_+^2Y_0^\prime\cdot\del_-^2Y_0^\prime 
-\del_-^2{Y_0^\mu}^\prime e_-\cdot\del_+^2Y_0^\prime-\del_+{Y_0^\mu}^\prime e_+\cdot\del_-^3Y_0^\prime \big) 
.\end{equation}
Substituting this in  \er{highT} one sees that to order $R^{-2}$ 
one gets the \emph{manifestly holomorphic} representation 
\begin{eqnarray}
\label{pscorrT}
T_{--}^\prime &=& -\frac{R}{a^2}e_-\cdot\del_-Y_0^\prime-\frac{1}{2a^2}\del_-Y_0^\prime\cdot
\del_-Y_0^\prime -\frac{\beta}{R}e_+\cdot\del_-^3Y_0^\prime \nn\\
&-&2\frac{\beta}{R^2}e_+\cdot\del_-^3Y_0^\prime e_+\cdot\del_-Y_0^\prime 
-2\frac{\beta}{R^2}(e_+\cdot\del_-^2Y_0^\prime)^2
,\end{eqnarray}
whence we obtain the Virasoro generators with higher-order corrections,
\begin{eqnarray}
\label{vircorr}
L_n^\prime &=& \frac{R}{a}e_-\cdot\alpha_n^\prime+\frac{1}{2}\sum_{m=-\infty}^\infty:\alpha_{n-m}^\prime\alpha_m^\prime:
+\frac{\beta_c}{2}\delta_{n} \\
&-&\frac{a \beta_c n^2}{R}e_n^\prime 
-\frac{\beta_c a^2 n^2}{R^2}\sum_{m=-\infty}^\infty :e_{n-m}^\prime e_m^\prime: \nn
,\end{eqnarray}
where $e_n^\prime\equiv e_+ \cdot \alpha_n^\prime$ (and likewise for ${\tilde L}_0$).

Therefore $L_0^\prime$ has no
corrections at either $R^{-1}$ or $R^{-2}$ order. 
It is important to note these corrected generators still satisfy \er{viralg} when
\er{ccr} is used.
It then follows that all the terms in the ground state energy and the
excited state energies, inclusive of the order-$R^{-3}$ term, are
identical to those in the Nambu-Goto theory.
\section{An all order analysis?}
\label{sec_all_analysis}
It is desirable to be able to carry out an \emph{all order} extension of the above mentioned results.
Such an analysis, for example, is crucial for a detailed comparison between effective string theories and
numerical simulations. A generalization of the formalism in \cite{ps} requires not only finding
the action to the desired order, but also determining the
appropriate conformal transformation laws under which it is invariant.  
Such an approach becomes tedious and
may be hard to make it systematic.
We have achieved this generalization
through a formulation wherein the transformation laws are independent
of the particular action chosen \cite{covariant}. We describe an important clue
that led to our formulation.

Recall that the PS proposal for the leading correction was based on 
the Liouville action(whose precursor is the Polyakov action for two dimensional quantum
gravity \cite{polya}) for subcritical strings
\begin{equation}
\label{liouville}
S_{Liou} = \frac{26-D}{48\pi}\int~d\tau^+~d\tau^-~~ \partial_+\phi\partial_-\phi
\end{equation}
They replaced the conformal factor
$e^\phi$ by the component
$L=\partial_+X\cdot\partial_-X$ (in the conformal gauge) of the induced
metric on the world sheet. They also replaced
$(26-D)/12$ by a parameter $\beta$ which was determined by
requiring the vanishing of the total central charge in all dimensions.\footnote{
They eventually found $\beta$ to be just the same as in the
Liouville theory. NDH thanks Prof Hikaru Kawai for explaining why this had to be so.}. 
These steps would have actually led to the total action
\begin{equation}
\label{stot}
S_{tot} = S_0 +S_{(2)}
\end{equation}
where
\begin{equation}
\label{action2}
S_{(2)} = \frac{\beta}{4\pi} \int \td\tau^+ \td\tau^- \bigg\{
\frac {\del_+(\del_+X\cdot\del_-X)\del_-(\del_+X\cdot\del_-X)}{(\del_+X\cdot\del_-X)^2}
\bigg\}
.\end{equation}
The important clue is that $S_{(2)}$ is \emph{exactly} invariant under
the conformal transformations
\begin{equation}
\label{urtrans}
\delta^{0}_\pm X^\mu = \epsilon^\pm(\tau^\pm)\del_\pm X^\mu
.\end{equation}
under which the leading action $S_0$ is also exactly invariant.
If we write $S_{(2)}$, as
\begin{equation}
\label{defL2}
S_{(2)} = \frac{\beta}{4\pi}\int \td\tau^+\td\tau^- {L}_2\quad\quad L_2 = \frac{\partial_-L\partial_+L}{L^2}
\end{equation}
it is easy to show that
\begin{equation}
\label{psanom}
\delta^{0}_- L_2 = \del_-(\epsilon^- L_2) + \del_-^2\epsilon^- \del_+ \ln L
\end{equation}
The first term is what one would have expected if $L_2$ had
transformed as a scalar density under \er{urtrans}.
The second term is a departure
from this. We shall explain this important point later. 
It should however be noted that the additional term can be rewritten as
\begin{equation}
\del_-^2\epsilon^- \del_+ \ln L = \del_+(\del_-^2\epsilon^- \ln L)
\end{equation}
ensuring the invariance of $S_{(2)}$. 



PS obtain their \er{psaction} by dropping terms proportional to $\partial_{+-}X^\mu$ (the EOM from the free action
$S_0$). The compensating infinitesimal field redefinition connects \er{urtrans} to \er{modtrans}.
The algebra of the PS transformations of \er{modtrans} is
\begin{equation}
[\delta^{PS}_-(\epsilon_1^-) , \delta^{PS}_-(\epsilon_2^-)] = \delta^{PS}_-(\epsilon_{12}^-) +\Pcm{O}(R^{-4})
,\end{equation}
where $\epsilon_{12}^- = \epsilon_1^-\del_-\epsilon_2^- - \epsilon_2^-\del_-\epsilon_1^-$. 
On the other hand, the algebra of the
transformations of \er{urtrans} is
\begin{equation}
[\delta^{0}_-(\epsilon_1^-) , \delta^{0}_-(\epsilon_2^-)] = \delta^{0}_-(\epsilon_{12}^-) 
.\end{equation}
Thus both generate the same group of symmetry transformations, namely
the conformal group. While the PS transformations realise this only
approximately, to $\Pcm{O}(R^{-4})$, 
the
transformations \eqref{urtrans} leaving $S_{2}$ invariant do so \emph{exactly}.

Therefore we have a candidate action $S_{(2)}$ that can in principle be expanded to arbitrary orders in $R^{-1}$,
and which is exactly conformally invariant. In this paper we use this particular action to study the issue of 
spectrum to all orders. We call $S_{(2)}$ an \emph{effective string action of the Polyakov-Liouville type} due of its obvious connection
to the Polyakov and Liouville actions.

It is very important to emphasize that such an analysis does not address the issue of the
spectrum to all orders of \emph{all} conformal effective string theories. The importance of our analysis is twofold;
on the one hand it would throw light on the possible nature of corrections to all orders, and on the other hand it
would address the issue of corrections to the spectrum of \emph{all} effective string theories to order $R^{-5}$
when combined with Drummond's result that there are no candidate effective actions at orders $R^{-4},R^{-5}$. We
carefully discuss the latter point in the section \ref{sec_discussion}.
\section{All order analysis of $S_{(2)}$}
\label{sec_liou_all}
Our analysis begins with a derivation of the EOM and $T_{--}$ corresponding to the total action $S_{tot}$ of \er{stot}.
We remind the reader that for generic actions these are given by
\begin{equation}
\label{defs}
E^\mu = \frac{\delta S}{\delta X^\mu}=0\quad\quad \delta S = \frac{1}{2\pi}\int \td\tau^+ \td\tau^-~\del_+\epsilon(\tau^-,\tau^+)~T_{--}
\end{equation}
the second part of which is obtained by treating the symmetry parameter $\epsilon(\tau^-)$ as depending on $\tau^+$ and deriving N\"other Theorem
as usual. The following relation between the \emph{off-shell} $T_{--}$ and $E^\mu$ is an important one:
\begin{equation}
\label{eomT}
\del_+~T_{--} = -2\pi E\cdot\del_-X
\end{equation}
The explicit expressions for the EOM and $T_{--}$ resulting from $S_{tot}$ are given by:
\begin{equation}
\label{fulleom}
4\pi E^\mu = -\frac{2}{a^2}\del_{+-}X^\mu -\beta[2Z~\del_{+-}X^\mu+\del_+Z~\del_-X^\mu+\del_-Z~\del_+X^\mu]
\end{equation}
\begin{equation}
\label{defZ}
Z = \frac{2}{L^3}~\del_+L\del_-L - 2\frac{\del_{+-}L}{L^2}
\end{equation}
\begin{equation}
\label{fullT}
T_{--} = \frac{1}{2a^2}\del_-X\cdot\del_-X-\frac{\beta}{2}[2\frac{\del_{--}L}{L}-3\frac{\del_-L\del_-L}{L^2}-\del_-X\cdot\del_-XZ]
\end{equation}
It is easily seen that \er{fullT} is not manifestly \emph{holomorphic}, a difficulty
we had encountered earlier in our $R^{-3}$ order analysis(see the discussion after \er{highT}). There we had integrated the EOM to the 
required accuracy and substituting
the solutions to the EOM had shown that the on-shell $T_{--}^\prime$ was manifestly holomorphic. Drummond had circumvented this difficulty
through an appropriate field redefinition. We adopt a different strategy now as integrating the full EOM to all orders is a very difficult
task and, as we shall see, unnecessary. 
We now introduce the decomposition 
\begin{equation}
\label{holosplit}
X^\mu = X^\mu_{cl}+F^\mu(\tau^+)+G^\mu(\tau^-) + H^\mu(\tau^+,\tau^-)
\end{equation}
where $F,G$ are anti-holomorphic and holomorphic functions respectively. It is important to note that $F,G$ can not have any polynomial
parts; this follows from the linearity of the boundary conditions. $H$ is purely \emph{non-holomorphic} and by construction
it does not have any purely holomorphic and anti-holomorphic parts. The advantage of introducing such a function is that solutions
to equations of the type
\begin{equation}
\label{Heqns}
\del_\pm H^\mu = \del_\pm W^\mu
\end{equation}
have the unique solution $H^\mu = W^\mu$. It is clear that \er{fulleom} can be recast in the form
\begin{equation}
\label{Heom}
\del_{+-}H^\mu = P^\mu_{+-}
\end{equation} 
On using the explicit form of $P^\mu_{+-}$ it is seen that $H$ is of order $R^{-3}$ and higher. The form of the full EOM given in \er{Heom} 
can be iteratively solved to give $H$ in terms of $F_+,G_-$ and their higher derivatives( here
$F_+ = \del_+F, G_-=\del_-G$. Additional $\pm$-indices indicate corresponding higher derivatives). Now using using \er{holosplit}, the
\emph{off-shell} $T_{--}$ of \er{fullT} can be split as
\begin{equation}
\label{Tsplit}
T_{--} = T_{--}^h + T_{--}^{nh}
\end{equation}
where $T_{--}^{nh}$ is \emph{purely non-holomorphic} in the sense that it has no holomorphic parts(it can however have anti-holomorphic pieces
and likewise $T_{++}$ can have holomorphic parts). Using \er{eomT} we can uniquely extract $T_{--}^{nh}$ and substituting that in \er{fullT} 
determine $T_{--}^h$ \emph{unambiguously}. The upshot of this is that the \emph{on-shell} $T_{--}$ can be obtained by simply setting 
$T_{--}^{nh} =0$. We simply quote the final \emph{all order} result:
\begin{equation}
\label{onshellT}
T_{--} = \frac{R}{a^2}~e_-\cdot G_-+\frac{1}{2a^2}G_-\cdot G_--\frac{\beta}{2}[2\frac{\del_{--}L_h}{L_h}-3\frac{\del_-L_h\del_-L_h}{L_h^2}]
\end{equation}
where
\begin{equation}
\label{holL}
L_h = -\frac{R^2}{2}+2R~e_+\cdot G_-
\end{equation}
is just the holomorphic part of 
\begin{equation}
\label{Lunprimed}
L = -\frac{R^2}{2}+R~e_+\cdot(G_-+H_-)+R~e_-\cdot(F_++H_+)+(F_++H_+)\cdot(G_-+H_-)
\end{equation}
We now expand \er{onshellT} as a power series in $R^{-1}$
\begin{eqnarray}
\label{seriesT}
T_{--} &=& \frac{R}{a^2}e_-\cdot G_-+\frac{1}{2a^2}G_-\cdot G_-+\frac{2\beta}{R}e_+\cdot G_{---}\nn\\
&+&\sum_{p=2}R^{-p}~\bigg[\beta_1^{(p)}\del_{--}(e_+\cdot G_-)^p+\beta_2^{(p)}~e_+\cdot G_{---}~(e_+\cdot G_-)^{p-1}\bigg ]
\end{eqnarray}
with
\begin{equation}
\label{seriesTcoef}
\beta_1^{(p)} = \frac{3\beta}{2}~\frac{2^p}{p}\quad\quad \beta_2^{(p)} = -\beta~2^{p-1}
\end{equation}
\section{Properties of a class of $T_{--}$}
\label{sec_propT}
The $T_{--}$ of \er{seriesT} is by no means the most general. It is sufficient for the purposes of this paper to consider
a \emph{somewhat larger} class of $T_{--}$, which however is also not the most general class:
\begin{eqnarray}
\label{largerT}
{\bar T}_{--} &=& \frac{R}{a^2}e_-\cdot {\bar G}_-+\frac{1}{2a^2}{\bar G}_-\cdot {\bar G}_-+\frac{{\bar\beta}^{(1)}}{R}e_+\cdot {\bar G}_{---}\nn\\
&+&\sum_{p=2}R^{-p}~\bigg[{\bar\beta}_1^{(p)}\del_{--}(e_+\cdot {\bar G}_-)^p+{\bar\beta}_2^{(p)}~e_+\cdot {\bar G}_{---}~(e_+\cdot {\bar G}_-)^{p-1}\bigg ]
\end{eqnarray}
Except for the part coming from the free theory, given by the first two terms, the stress tensors of \er{seriesT} and 
\er{largerT} are entirely made up of $e_+^\mu$ and $-$derivatives of $G_-$; since each term has to
have exactly \emph{two} net $-$indices, the most general terms of this type are the ones shown in \er{largerT}. The $T_{--}$ associated 
with our Polyakov-Liouville type effective theory is given by $\beta^{(1)}=2\beta$ and the remaining coefficients as in \er{seriesTcoef}.
Now we prove some important properties of stress tensors belonging to this class. 
\subsection{Field redefinitions maintaining the class}
\label{sec_redef_same}
We consider the class of \emph{small field redefinitions} which too depend on $e_+^\mu$ only. Now each term must have only \emph{one}
net $-$index. It is easy to see that the most general field redefinition of that type is given by
\begin{equation}
\label{redefgen}
G_-^\prime = G_-+e_+^\mu~\frac{a^2}{R^2}\bigg[\gamma^{(1)}~e_+\cdot G_{---}+\sum_{p=2}~R^{-p+1}\bigg \{\gamma_1^{(p)}~
\del_{--}~(e_+\cdot G_-)^p + \gamma_2^{(p)}~e_+\cdot G_{---}~(e_+\cdot G_-)^{p-1}\bigg \}\bigg ]
\end{equation}
An important consequence is $e_+\cdot G_-^\prime = e_+\cdot G_-$ because $e_+\cdot e_+ = 0$. This means \er{redefgen}
is easily \emph{inverted}:
\begin{equation}
\label{invredef}
G_- = G_-^\prime-e_+^\mu~\frac{a^2}{R^2}\bigg[\gamma^{(1)}~e_+\cdot G_{---}^\prime+\sum_{p=2}~R^{-p+1}\bigg \{\gamma_1^{(p)}~
\del_{--}~(e_+\cdot G_-^\prime)^p + \gamma_2^{(p)}~e_+\cdot G_{---}^\prime~(e_+\cdot G_-^\prime)^{p-1}\bigg \}\bigg ]
\end{equation}
Also the transformations of \er{redefgen} form an \emph{Abelian Group}. It can easily be shown that under a transformation given
by \er{invredef}, a ${\bar T}_{--}$ of
the type in \er{largerT} parametrised by $\{\beta_1^{(k)},\beta_2^{(k)}\}$(we shall henceforth follow the convention that
$\beta_1^{(1)}=\beta_2^{(2)}=\frac{\beta^{(1)}}{2},\gamma_1^{(1)}=\gamma_2^{(1)}=\frac{\gamma^{(1)}}{2}$) gets mapped to a ${\bar T}_{--}^\prime$ with parameters $\{{\beta_1^{(k)}}^\prime,
{\beta_2^{(k)}}^\prime\}$ given by
\begin{eqnarray}
\label{paramredef}
{\beta_1^{(k)}}^\prime&=&\beta_1^{(k)}-\frac{\gamma_1^{(k)}}{2}+\frac{k-2}{k}\gamma_1^{(k-1)}\nn\\
{\beta_2^{(k)}}^\prime&=&\beta_2^{(k)}-\frac{\gamma_2^{(k)}}{2}+\gamma_1^{(k-1)}+\gamma_2^{(k-1)}
\end{eqnarray}
In \er{paramredef} $k \ge 1$ and $\gamma_1^{(0)}=\gamma_2^{(0)}$. The $\gamma$'s can be solved for \emph{uniquely}:
\begin{eqnarray}
\label{gammasol}
\gamma_1^{(1)}&=&\gamma_2^{(1)}=2(\beta_1^{(1)}-{\beta_1^{(1)}}^\prime)\nn\\
\gamma_1^{(2)}&=& 2(\beta_1^{(2)}-{\beta_1^{(2)}}^\prime)\quad\quad \gamma_2^{(2)}=2(\beta_2^{(2)}-{\beta_2^{(2)}}^\prime)
+8(\beta_2^{(1)}-{\beta_2^{(1)}}^\prime)
\end{eqnarray}
and for $k\ge 2$,
\begin{eqnarray}
\label{gammasol-k}
\gamma_1^{(k)}&=&\frac{2^k}{k(k-1)}\sum_{i=2}^k2^{1-i}i(i-1)(\beta_1^{(i)}-{\beta_1^{(i)}}^\prime)\nn\\
\gamma_2^{(k)}&=&2^{(k-1)}~\gamma_2^{(1)}+2^k\sum_{j=2}^k 2^{1-j}(\beta_2^{(j)}-{\beta_2^{(j)}}^\prime +\gamma_1^{(j-1)})
\end{eqnarray}
\section{Virasoro Generators and Spectrum}
\label{spectrumall}
With the mode expansion $G_-^\mu=a\sum~\alpha_m^\mu~e^{-im\tau^-}$ the oscillator representation for the Virasoro generators 
$L_m$ corresponding to \er{seriesT} is given, after normal ordering, by
\begin{eqnarray}
\label{allVir}
L_m &=& \frac{R}{a}~e_-\cdot\alpha_m+\frac{1}{2}\sum_n~:\alpha_{m-n}\cdot\alpha_n:-\frac{2a\beta m^2}{R}e_+\cdot \alpha_m+\frac{\beta}{2}~\delta_{m,0}\nn\\
    &-&\frac{\beta}{2}\sum_{p=2}\frac{1}{p}(\frac{2a}{R})^p\sum_{n_1..n_p}~\prod_{i=1}^pe_+\cdot\alpha_{n_i}~\delta_{\sum_j~n_j,m}~
\{3m^2-\sum_l~n_l^2\}
\end{eqnarray}
A suitable commutation relation has to be found between the quantum operators $\{\alpha_m^\mu\}$ that would result in
Virasoro algebra for $L_m$. Though $G^\mu$ satisfies the
\emph{free} equation, it would be incorrect to invoke the canonical commutation relations of \er{ccr}. In fact it is easy, though
tedious, to verify that they do not reproduce the correct Virasoro Algebra. In fact, any holomorphic field redefinition of $G^\mu$
also satisfies the free EOM. Therefore the best we can expect for $[\alpha_m^\mu,\alpha_{m^\prime}^\nu]$ is one that under a suitable
field redefinition would take the form of \er{ccr}. After some algebra it can be shown that the required commutator is
\begin{eqnarray}
\label{allalphacr}
[\alpha_m^\mu,\alpha_{m^\prime}^\nu] &=& m\eta^{\mu\nu}\delta_{m+m^\prime,0}
+e_+^\mu e_+^\nu~\frac{2\beta a^2}{R^2}mm^\prime(m-m^\prime)
\bigg[\delta_{m+m^\prime,0}\nn\\
&+&\sum_{k=2}k(\frac{2a}{R})^{k-1}\sum_{\{n_i\}}\delta_{\sum n_i,m+m^\prime}\prod_{l} e_+\cdot 
\alpha_{n_l}\bigg]
\end{eqnarray}
This precisely reproduces \er{viralg} which also shows that our extended theory is also quantum mechanically consistent in
all dimensions after the choice of \er{betacrit} is made.

The quantisation procedure outlined earlier \emph{implicitly} relies on certain consistency conditions on the oscillator algebra;
these are $[L_m,\alpha_0^\mu]=0$ for $m\ge 0$. A \emph{sufficient} condition is $[\alpha_0^\mu,\alpha_m^\mu]=0$ for $m\ge 0$.
Relations of \er{ccr} indeed satisfy these. There is no guarantee that general ccr's even when they are reducible to \er{ccr}
under field redefinitions would satisfy these. However, the commutators of \er{allalphacr} do satisfy these! Before proceeding we 
emphasize two important properties: $[e_+\cdot\alpha_m,e_+\cdot\alpha_{m^\prime}]=0$(for all $m,m^\prime$) and 
$[\alpha_m^\mu,e_+\cdot\alpha_n]=me_+^\mu\delta_{m,-m^\prime}$. Because of the first of these relations the \emph{factor ordering} 
issues in the new $L_m$'s are the same as in the free theory.

It also means that in the field redefinitions of \er{redefgen}, now viewed as redefinitions among oscillator operators, there are 
no factor ordering issues. This has the deeper implication that the equivalence under field redfinitions theorem of classical field
theory can now be taken over for the class of quantum field theories considered here. In particular, the entire class of ${\bar T}_{--}$
in \er{largerT} represents just one physical theory because of the results of section \ref{sec_propT}. 

One immediately sees from \er{allVir} that the ground state energy to all orders in $R^{-1}$ is \emph{exactly} the same as in free
bosonic string theory. The analysis of the spectrum of excited states based on \er{allVir} and \er{allalphacr} is however more involved. In
principle, with such generalized commutation relations even the orthogonality of different multiparticle states is no longer guaranteed.
However in this particular case such problems are not there though sector-wise transformations in target space have to be carried
out to realise full orthonormality. 
We approach the problem of the spectrum to all orders somewhat differently.

The Virasoro generators corresponding to \er{largerT} can be written down:
\begin{eqnarray}
\label{largerVir}
{\bar L}_m &=& \frac{R}{a}~e_-\cdot{\bar \alpha}_m+\frac{1}{2}\sum_n~:{\bar \alpha}_{m-n}\cdot{\bar \alpha}_n:-\frac{a{\bar \beta}^{(1)} m^2}{R}e_+\cdot {\bar \alpha}_m+\frac{{\bar \beta}^{(1)}}{2}~\delta_{m,0}\nn\\
    &-&\sum_{p=2}(\frac{a}{R})^p\sum_{n_1..n_p}~\prod_{i=1}^pe_+\cdot{\bar \alpha}_{n_i}~\delta_{\sum_j~n_j,m}~
\{{\bar \beta}^{(p)}_1~m^2+\frac{{\bar \beta}_2^{(p)}}{p}\sum_l~n_l^2\}
\end{eqnarray}
We now ask how the parameters $\{{\bar \beta}^{(1)},{\bar\beta}_1^{(p)},{\bar\beta}_2^{(p)}\}$ of \er{largerVir}
get restricted if we require that one gets the Virasoro Algebra of \er{viralg} when \er{ccr} is used for
${\bar\alpha}_m^\mu$:
\begin{equation}
\label{largerccr} 
[{\bar\alpha}_m^\mu,{\bar\alpha}_n^\nu]~= m\eta^{\mu\nu}\delta_{m,-n}
\end{equation}
We have shown that all the parameters of \er{largerVir} get determined according to
\begin{equation}
\label{largerps}
{\bar\beta}^{(1)} = \beta\quad\quad{\bar\beta}_1^{(p)}=\beta\frac{2^{p-1}}{p}\quad\quad {\bar\beta}_2^{(p)}=0
\end{equation}
It is worth recording the associated $L_m, T_{--}$:
\begin{eqnarray}
\label{allpsVir}
 L_m^\prime &=& \frac{R}{a}~e_-\cdot \alpha^\prime_m+\frac{1}{2}\sum_n~: \alpha^\prime_{m-n}\cdot \alpha^\prime_n:
+\frac{ \beta}{2}~\delta_{m,0}\nn\\
    &-&
\frac{a \beta m^2}{R}e_ +\cdot  \alpha^\prime_m
-\frac{\beta m^2}{2}\sum_{p=2}\frac{1}{p}(\frac{2a}{R})^p\sum_{n_1..n_p}~\prod_{i=1}^pe_+\cdot \alpha^\prime_{n_i}~\delta_{\sum_j~n_j,m}~
\end{eqnarray}
\begin{equation}
\label{allpsT}
T_{--}^\prime = \frac{R}{a^2}e_-\cdot{G_-}^\prime+\frac{1}{2a^2}{G_-}^\prime\cdot{G_-}^\prime -
\frac{\beta}{2}~\del_{--}~\ln(1-2\frac{e_+\cdot G_-^\prime}{R})
\end{equation}
These expressions when expanded to order $R^{-2}$ precisely reproduce \er{vircorr} and \er{pscorrT}. From the results of section 
\ref{sec_propT} we can read out the parameters in \er{redefgen} that corresponds to the unique field redefinition mapping \er{onshellT}
to \er{allpsT}:
\begin{equation}
\label{2tops}
\gamma^{(1)} = 2~\beta\quad\quad \gamma_1^{(k)} = 2^k~\beta\quad\quad \gamma_2^{(k)}=0
\end{equation}
Not surprisingly, the same redefinition also maps \er{allalphacr} to \er{ccr}!

One sees immediately from \er{allpsVir} that the operators $L_0^\prime( {\tilde L}_0^\prime)$ are \emph{identical} to their free
field expressions. With the oscillator algebra also remaining of the same form as in \er{ccr}, the spectrum of all states in the
Polyakov-Liouville type effective theory, to all orders in $R^{-1}$, is exactly the same as that of the free bosonic string theory.
\section{Discussion and Conclusions}
\label{sec_discussion}
In this paper we have first constructed an effective string theory, called the \emph{Polyakov-Liouville Type} by us, which is exactly
conformal invariant to all orders in the small parameter $R^{-1}$. This is a significant extension of the original work by Polchinski and
Strominger, and later, by Drummond and us. The form of the transformation laws is as in the free theory. We are able to 
carry out the analysis of 
$T_{--}$, the EOM, the Virasoro Generators and their associated oscillator algebras and the spectrum of the theory to all orders. 
We have shown that to all orders the spectrum of the Polyakov-Liouville type theory is the same as that of the 
free bosonic string theory. It, like the original PS-theory, is quantum mechanically consistent in all dimensions.

It should be stressed that we have not analysed the spectrum of arbitrary classes of effective string theories of the PS-type.
However, Drummond \cite{drumorig} has argued that there are no additional candidate actions even at $R^{-4},R^{-5}$ levels. If
so, the results of our paper would imply that the spectrum of \emph{all} conformal effective string theories would be the same
as that of free bosonic string theory(Nambu-Goto action) to these levels. Aharony and Karzbrun \cite{aharony} have claimed that
this is so for $d=3$, but for higher dimensions they claim that there are potential corrections. The situation is somewhat reminiscent
of the analysis at $R^{-3}$ order; while the PS-approach suitably generalised as in \cite{drumorig,ouruniv} had shown the absence of
corrections for all dimensions, according to the L\"uscher-Weisz analysis \cite{lwrecent} this was so only for $d=3$. These differences
need to be narrowed in future analysis. The covariant calculus developed by us \cite{covariant} would be the way to bridge the Polchinski-Strominger and L\"uscher-Weisz approaches. Much work needs to be done along these directions.

On the other
hand, Drummond had invoked dropping terms proportional to leading order EOM. If such terms are dropped by performing suitable field
redefinitions, issues that had been raised in \cite{field} for \emph{generic} field redefinitions will have to be carefully analysed.
Of course, the kind of field redefinitions implicitly used by Drummond may also turn out to be \emph{benign} as the field redefinitions
used in the current work. It could also be that certain terms proportional to $\del_{+-} X$ can be dropped for other reasons without doing any field redefinitions at all.

\acknowledgments{Both NDH and YB gratefully acknowledge the DAE Raja Ramanna Fellowship Scheme under which this work was carried
out, and CHEP, Indian Institute of Science, Bangalore, for their facilities. NDH wishes to thank Hikaru Kawai, T. Yukawa, Ashoke 
Sen, Vikram Vyas and Gunnar Bali for many valuable discussions. We thank all members of the Working Group on QCD-Strings at the International Workshop 
\emph{StrongFrontier 2009} for stimulating discussions. 

\end{document}